\documentclass[twocolumn,showpacs,amsmath,amssymb]{revtex4}
\input epsf
\usepackage{appendix}
\usepackage{rotating}
\begin{document}
\title {Anisotropic Coherence properties in a trapped quasi2D dipolar gas}
\author{Christopher Ticknor}
\affiliation{Theoretical Division, Los Alamos National Laboratory, Los Alamos, 
New Mexico 87545, USA}
\date{\today}
\begin{abstract}
We consider a trapped quasi2D dipolar Bose Einstein condensate (q2D DBEC) 
with a polarization tilted into the plane of motion. 
We show that by tilting the polarization axis, 
the coherence properties are anisotropic.
Such a system will have density fluctuations whose amplitude
depend on their trap location.
Additionally, interference contrast will also be anisotropic
despite an isotropic density profile.
The anisotropy is related to a roton like mode that becomes unstable
and supports local collapse along the polarization axis.
\end{abstract} 
\maketitle
Coherence properties underlie collective quantum behavior. 
An example of a need to further understand 
coherence is in high $T_c$ superconductors \cite{phaseTC},
where the origin of the phase coherence leading to superconducting behavior
is still mysterious.  Such solid state systems where this occurs
are complex, making the study of coherence properties a challenge.
One avenue to study coherence of quantum systems more clearly is
in ultracold gases, which have offered direct insight and unprecedented 
control to the study of quantum collective behavior.
For example, two macroscopic Bose Einstein condensates (BECs) were
interfered displaying phase coherence \cite{int}.
More refined experiments have directly detected higher order correlations
\cite{correlations} which cannot be easily accessed by electronic systems.
Additionally, reduced dimensional ultracold gases have been the focus of 
intense interest because fluctuations are enhanced in such geometries.
Phase coherence has been studied at the 
Berezinskii Kosterlitz Thouless (BKT) transition \cite{BKT,BKTrev}
where two independent q2D
ultracold gases were interfered, revealing thermally activated phase defects
are the origin of decoherence \cite{zoran}.
The BKT transition is a transition where the phase coherence changes 
character, and insights into controlling such coherence properties are 
of general importance.
In this paper we show the coherence properties of a trapped dipolar
gas can be controlled.

The progress of experiments to probe ultracold gases is remarkable.
Recent advances in imaging q2D ultracold gases have been used 
to detect density fluctuations \cite{chicago,jila,ENS}. Additionally,
ultracold atomic systems in one dimensional systems have also amazingly 
been used to study quantum fluctuations directly \cite{armijo2} and other 
coherence properties \cite{armijo1,manz,imam,kinoshita}.
For these examples of ultracold gases, the interaction between the 
constituents are short range.  There are ultracold dipolar systems which have 
anisotropic, long range interactions, which offer an entirely  
new avenue to study quantum correlations and collective behavior. 
Amazing experimental progress has been made with the strongly magnetic atoms:
Chromium (Cr), Dysprosium (Dy), and Erbium (Er).  All three have been Bose 
condensed and displayed strong dipolar effects \cite{Cr,Dy,Er}.
The study of dipoles in reduced geometries have begun with Cr, where it
has been put into a one dimensional lattice, forming 
a series of coupled q2D dipolar BECs \cite{layer}. 
Additionally, polar molecules have been loaded into q2D geometries to control
the rate of their chemical reaction \cite{rbk}.

Correlations of ultracold dipolar gases have been studied with 
a perpendicular polarization for weakly interacting gases
near the roton instability \cite{klaw,sykes,blair}
and strongly interacting gases, e.g. \cite{buchler}.
Using the direction of the polarization axis as a means to control the 
collective behavior and coherence properties, 
a q2D dipolar gas has been studied in the many body dipolar systems at the 
mean field level \cite{nath,aniso} and for 2D dipolar scattering \cite{tilt}.
Additional theories have investigated a tilted polarization axis in a 
q2D dipolar fermi gas \cite{tilt2}.

In this work, we go beyond mean field theory and examine the 
correlations of a trapped q2D DBEC at finite temperature with a partially tilted
polarization into the plane of motion.
This situation leads to anisotropic coherence properties, in which
the gas maintains coherence along the direction of 
polarization, but more rapidly loses coherence perpendicular 
to the polarization direction.
This anisotropy is evident in both the phase and density-density coherence 
properties.  This emerges because of a roton-like mode in the 
excitation spectrum and is strongly anisotropic in character.
We consider the implications of this theory through interference experiments
and density fluctuations, both of which will exhibit anisotropic 
correlated behavior.
We also calculate the compressibility of the dipolar gas.

To study the finite temperature coherence properties, we employ the Hartree Fock Bogoliubov 
method within the Popov approximation  (HFBP) with nonlocal 
interactions \cite{griffin,CThfb}.
The HFB breaks the wavefunction into a condensate and thermal component:
$\hat\Psi=[\sqrt{N_0}\phi_0(\vec\rho)+\hat\theta(\vec\rho)]$ where we have 
replaced $\hat a_0\rightarrow\sqrt{N_0}$.  We use the Bogoliubov transformation:
$\hat\theta(\vec\rho)=\sum_\gamma [u_\gamma(\vec\rho)\hat a_\gamma e^{-i\omega_\gamma t}
-v_\gamma^*(\vec\rho)\hat a^*_\gamma e^{i\omega_\gamma t}]$ where $\hat a_\gamma$ ($\hat a^*_\gamma$) is the bosonic annihilation (creation) operator for the $\gamma^{th}$ quasiparticle. 
The HFBP solves a non-local, generalized Gross-Pitaevskii equation and 
a nonlocal Bogoliubov de Gennes set of equations from which the 
eigenvalues and vectors (quasiparticles) are 
self-consistently obtained \cite{CThfb}.

To evaluate the interaction, we assume a dipole moment of the form: 
$\vec d=d[\hat x\sin(\alpha)+\hat z\cos(\alpha)]$ where $\alpha$ is the angle
between $\hat z$ and $\hat d$, and we assume a single transverse 
Gaussian wavefunction of width $l_z$ is occupied in the $z$ direction,
where $l_i=\sqrt{\hbar /\omega_i m}$ and $\omega_i$ is the trapping frequency.
This requires $\mu,T\ll\hbar\omega_z$, and this leads to 
a simple form of the interaction. We use
an isotropic trap ($x,y$) with $l_z/l_\rho=0.15$
($\omega_z/\omega_\rho$=44.4), and an interaction strength of 
$g_d=0.025$  and $g=0$, where $g_d=d^2/\sqrt{2\pi}l_z$ and
$g=\sqrt{8 \pi}a/l_z$ with $a$ being the 3D s-wave scattering length ($a\ll l_z$). 
The polarization angle is set to $\alpha/\pi=0.25$, this is beyond the critical angle. 
The critical angle is  when the interaction is zero along the $x$ axis, so
$\sin(\alpha_c)=1/\sqrt{3}$.
For $\alpha>\alpha_c\sim0.2\pi$ there is an attractive region to the interaction.
This value of $\alpha$ is picked to optimize the collapse with respect to
particle number. If it were steeper, the critical number would be less, 
and the depletion would not be large enough to observe the desired behavior.
In addition, if $\alpha$ were smaller then roton the mode would require many 
more particles to collapse, and the collapses would be more 3D in nature.  
For $\alpha/\pi=0.25$, the system is well described by the q2D 
formalism.
For these parameters, the trapping potential for 
Dy [Er] is $(\omega_\rho,\omega_z)/2\pi$= (12,533) [(48,2133)] Hz. 
In a trapped ideal 2D gas, Bose condensation can occur \cite{trap}, and the 
the critical temperature is at $T_C/\hbar\omega=\sqrt{6N}/\pi\sim0.78\sqrt{N}$.
This is for the pure 2D trapped gas, we only use it as a reference 
temperature.  Throughout this paper we work in trap units.
It is important to mention that the examples given here are weakly 
interacting; the HFBP method cannot accurately handle strongly interacting 
dipolar gases ($\mu<10\hbar\omega_\rho$) \cite{CThfb}.  
This criteria is from the deviation of the Kohn mode, or slosh mode, from 1$\hbar\omega_\rho$.
If the interaction is too strong then the static thermal cloud  alters the 
slosh mode of the condensate; this is a numerical signature that the HFBP has broken down.
This restricted the work from
entering the Thomas Fermi regime where $l_\rho>\xi$, where $\xi$ is the healing length 
($\sqrt{m\mu/\hbar^2}$).  Because of this, trapping effects are always important in this
study.

Once $g_d$, $g$, $l_z$, $T$, $\alpha$ and the condensate number, $N_0$, are 
selected, the HFBP calculation finds: the number of thermal atoms, 
$\tilde N$; the chemical potential, $\mu$; the condensate wavefunction, 
$\phi_0$; and the quasiparticle wavefunctions, $u_\gamma$ and $v_\gamma$.
From these wavefunctions, we construct the correlation functions:
the nonlocal thermal correlation function, 
$\tilde n(\vec \rho,\vec \rho^\prime)=\sum_\gamma [u^*_\gamma(\vec\rho) u_\gamma(\vec\rho^\prime)+v^*_\gamma(\vec\rho) v_\gamma(\vec\rho^\prime)
]N_{\gamma}+v^*_\gamma(\vec\rho) v_\gamma(\vec\rho^\prime)$ 
with $N_\gamma$ being the 
Bose-Einstein occupation, and the condensate correlation function, 
$n_0(\vec \rho,\vec \rho^\prime)$, given by $N_0\phi^*_0(x)\phi_0(x^\prime)$.  
The total local density is $n(\vec\rho)=n_0(\vec \rho)+\tilde n(\vec \rho)$ 
where e.g. $n_0(\vec\rho)=n_0(\vec\rho,\vec \rho)$.
The HFBP method uses these correlation functions in the self-consistent 
calculation, and they can be related to the more common $g_1$ and $g_2$
correlation functions \cite{phase}:
\begin{eqnarray}
g_1(\vec \rho;\vec \rho^\prime)&&={\langle \Psi^*(\vec\rho)\Psi(\vec\rho^\prime)\rangle\over \sqrt{n(\vec \rho)n(\vec \rho^\prime)}} =
{n_0(\vec \rho,\vec \rho^\prime)+\tilde n(\vec \rho,\vec \rho^\prime)\over\sqrt{n(\vec \rho)n(\vec \rho^\prime)}},\\ 
g_2(\vec \rho;\vec \rho^\prime)&&={\langle \Psi^*(\vec\rho)\Psi^*(\vec\rho^\prime) \Psi(\vec\rho)\Psi(\vec\rho^\prime)\rangle\over {n(\vec \rho)n(\vec \rho^\prime)}}
\\&&=1+g_1(\vec \rho,\vec \rho^\prime)^2-{n_0(\vec \rho,\vec \rho^\prime)^2\over n(\vec \rho)n(\vec \rho^\prime)}.\nonumber
\end{eqnarray}
$g_1$ details phase correlations, for example it determines the fringe 
contrast if a BEC were split and interfered. 
$g_2$ is the density-density correlation function, and is related to 
density fluctuations and the static structure factor \cite{zam}.
 
We study a system with a temperature of $T/T_C=0.5$.
At this temperature, the condensate fraction is
$N_0/N>0.57$ for all $N$ considered.  This is well within the region
of validity for the HFBP.
This temperature thermally populates the quasiparticles,
so that the anisotropy in the coherence properties is significant.
Furthermore, this temperature makes the density more isotropic.
Values of $T/T_C$ between $ 0.4$ and $0.6$ produces similar results.
Raising the temperature would lead to populating more thermal modes,
and would make correlations isotropic, reflecting the nature of the 
trap ($T/T_C\sim1$).   For lower temperatures, the correlations are determined 
by the condensate's properties alone, as the depletion and thermal population 
become negligible.  
A pure condensate has total phase coherence, it is the thermal fraction 
that leads to reduced coherence.

\begin{figure}
\includegraphics[width=80mm]{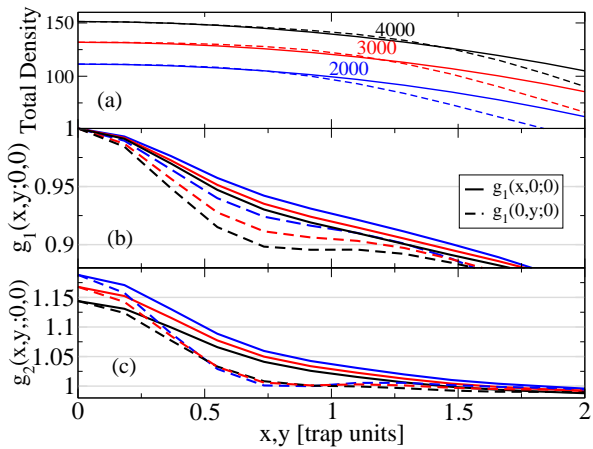}
\caption{(a) The total density, 
(b) $g_1(x,0;0,0)$ (solid line) and $g_1(0,y;0,0)$ (dashed), and 
(c) $g_2(x,0;0,0)$ (solid line) and $g_2(0,y;0,0)$ (dashed)
 at $T/T_C$=0.50 and for $N$ =2000 (blue), 3000 (red), and 4000 (black)
from top to bottom.
The total density is isotropic in the center and yet the correlation 
function is significantly anisotropic at short range.
}\label{g1xy}
\end{figure}

We now look at the correlation functions, $g_1(\vec\rho;0)$ and 
$g_2(\vec\rho;0)$.  These quantify the coherence of the system
from the origin to $\vec\rho$.
To illustrate the anisotropic coherence of the gas, we plot them along the
polarization axis and in the perpendicular direction in 
figure \ref{g1xy}.

Fig. \ref{g1xy} (a) shows the total density and Fig. \ref{g1xy} (b) shows
$g_1(\vec \rho;0)$ along the $x$ (solid line) and $y$ (dashed) axes. 
The important feature is that $g_1$ quickly decreases in the $y$ direction, 
but only steadily decreases its value in the $x$ direction or polarization 
direction.
We have shown the system for three different number of total particles
2000 (blue), 3000 (red), and 4000 (black), from top to bottom at the
temperature $T/T_C=0.5$.
Varying the number is like varying the interaction strength.
The density is isotropic near the center of the trap, even 
when the $g_1$ is anisotropic.  Near the edge of the trap the 
total density becomes anisotropic and is more extended in the $x$ axis.

A recent Bragg spectroscopy experiment observed an
anisotropic speed of sound for a 3D dipolar gas \cite{bragg}.
A similar experiment for a q2D DBEC could be used to probe these
predicted anisotropic coherence properties.  If a DBEC were split
into two equal clouds such that one is moved by $\vec a$ with
a relative phase of $S(\vec\rho)$, then the total density would be
$n_{}(\vec\rho)=0.5(n(\vec\rho)+n(\vec\rho+a))+\sqrt{n(\vec\rho)n(\vec\rho+\vec a)}g_1(\vec\rho,\vec\rho+\vec a)\cos[S(\vec\rho)]$.
The fringe contrast will depend on the angle between $\vec a$ and $\vec d$
because of the anisotropy in $g_1(\vec\rho,\vec\rho+\vec a)$.
To maximize the difference in fringe contrast, the phase difference should be
$\pi/2$ when $\vec a$ is near 3/4 $l_\rho$. This signal is only a
few percent and would be challenging to measure.

We now move on to the density-density correlation function,  
$g_2(\vec\rho;0)$.  In Fig. \ref{g1xy} (c), $g_2$ this is plotted
along the $x$ (solid line) and $y$ (dashed) axis.
The results are similar to those of $g_1$, 
the extent of the coherence is further in the $x$ direction 
than the $y$. The anisotropy is more significant in this correlation 
function. The fact that the $g_2$ goes below one is a trap effect and is 
most significant when correlating with the trap center \cite{manz,imam}.
This correlation function is important for density fluctuations, this will
be discussed below.

\begin{figure}
\includegraphics[width=85mm]{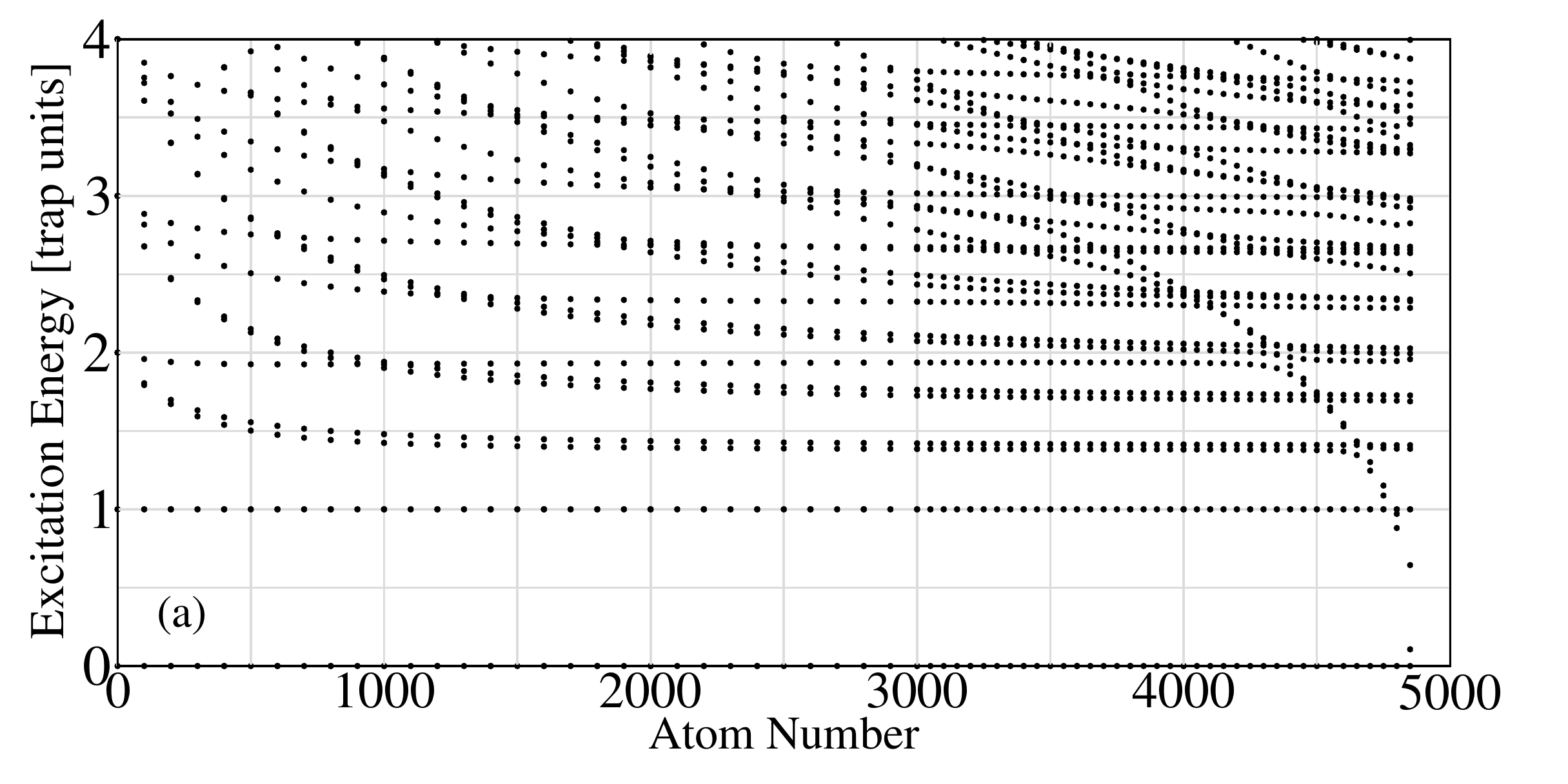}
\includegraphics[width=80mm]{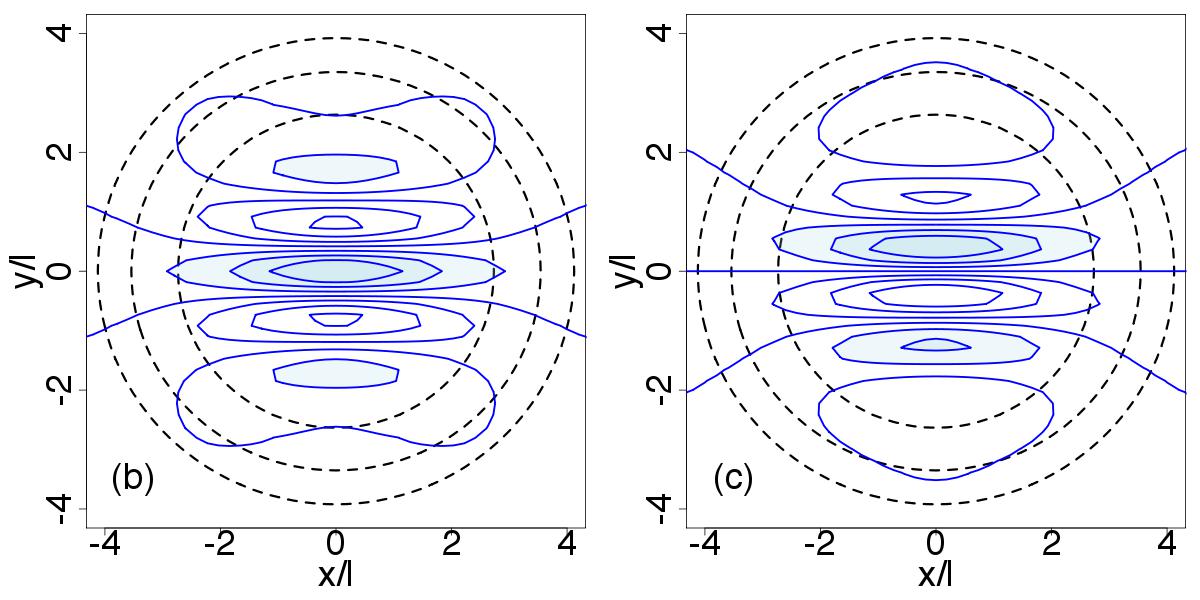}
\caption{(a) The Bogoliubov de Gennes spectrum is shown as a function of 
particle number for $g_d=0.025$ for $\alpha/\pi=0.25$.
(b,c) A contour plot of the roton quasiparticle modes ($u_\gamma$) 
which go soft for a dipolar system with tilted polarization axis are also 
shown.  The dashed black line is $\sqrt{n_0}$ and for $u_\gamma$ the  
shaded regions are less than zero. The contours at 0.25, 0.5, and 0.75 the 
maximum values of the individual wavefunctions.}\label{spectrum}
\end{figure}

To understand the nature of the anisotropy we look at the 
Bogoliubov de Gennes excitation spectrum as a function 
of particle number, this is shown in Fig. \ref{spectrum} (a).  
The spectrum goes soft or an eigenvalue becomes complex implying
the system is unstable 
when there are more than 4850 particles. 
Additionally, we show contour plots of the wavefunctions of the 
two modes which go soft ($u_\gamma$) in Fig. \ref{spectrum} (b,c), and 
the dashed black lines are $\sqrt{n_0}$. The contours at 0.25, 0.5, 
and 0.75 of the maximum values of the individual wavefunctions, 
and for $u_\gamma$ the shaded regions signify a region less than zero.
These quasiparticle modes start with an energy at 
or above $4\hbar \omega_\rho$. 
Then as the particle number is increased, their energy decreases.
In a homogeneous q2D DBEC, the roton is a local minimum in the 
excitation spectrum, which occurs near the $k_\rho l_z\sim1$ \cite{roton1}.
If the spectrum goes soft, the system 
locally collapses into plane waves which form dense stripes along
the polarization axis. 

The Bogoliubov de Gennes excitation spectrum has been analyzed in detail 
in Ref. \cite{CThfb} for a dipolar gas with $\alpha=0$ or a
cylindrically symmetric system.  In that case there are degenerate modes with 
azimuthal symmetry $\pm m$.  In the present case this degeneracy has been 
broken by the tilted polarization axis, leading to slightly different
energies depending on the excitation shape relative to the $x$ axis.  This 
leads to band like structures forming in Fig. \ref{spectrum} (a).

The mode in \ref{spectrum} (b) makes a significant contribution 
to the correlation function for two reasons:
first, it is a low energy  mode, and is more thermally populated.
Second, the correlation function, $g_i(\vec\rho:0)$,
has a factor of $u_\gamma(0)$ and $v_\gamma(0)$ in it.
Only a few low energy modes have a significant contribution
at the origin. The breathing mode just below 2$\hbar\omega$
and the roton mode in \ref{spectrum} (b) are two low energy modes with a maximum
in the center of the trap.  Other low energy modes have
small or zero amplitude in the center of the cloud because symmetry.

In the trapped case, the finite extent of the gas alters the nature
of the ``roton'' mode  \cite{roton,Wilson,Wilson2}. 
For $\alpha=0$, the gas become 3D in nature, $\mu\sim\hbar\omega_z$ \cite{blair}.
When the gas becomes unstable, it locally collapses and forms regions of high density.
A good demonstration of this is given in Ref. \cite{parker}.
For $\alpha/\pi\sim0.25$, the stability of the system displays an interplay of the interaction 
and the trap geometry, while maintaining $\mu\ll\hbar\omega_z$.
The collapse is local and forms a series density regions parallel to the polarization.
In Fig. \ref{spectrum} (b,c), we see this character in
the quasiparticles, which have an extended amplitude along the 
polarization axis and a series of nodes in the perpendicular direction.
It is this character of the quasiparticles that leads to anisotropic 
character to the correlations.  The nodes in the roton modes lead 
to destructive interference and reduce correlation in that direction.

\begin{figure}
\includegraphics[width=80mm]{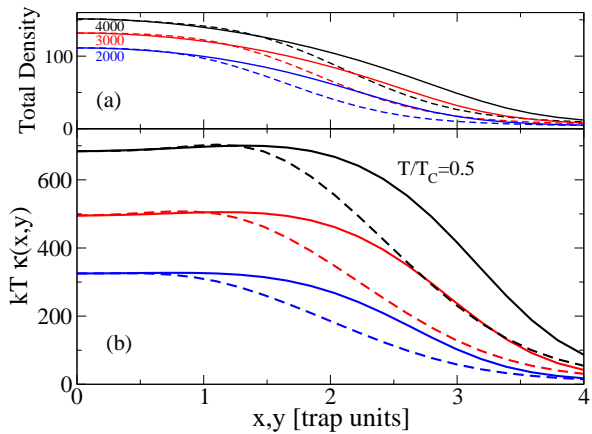}
\caption{(a) The total density (b) the compressibility, $kT \kappa(\vec \rho)$, for the dipolar gas along the $x$ (solid) and $y$ (dashed) axis
for particles number 2000 (blue), 3000 (red), and 4000 (black).
}\label{compress}
\end{figure}

To further study the implications of the anisotropic correlations, we consider 
the density fluctuations of the gas by calculating the compressibility:
$kT \kappa(\vec \rho)=$
$\langle \int d \vec\rho^\prime \delta n(\vec\rho)\delta n(\vec\rho^\prime)\rangle$
=$\int d \vec \rho ^\prime n(\vec \rho)n(\vec \rho^\prime)g_2(\vec \rho;\vec \rho^\prime)-n(\vec\rho)(N-1)$
where $\delta n(\vec\rho)=\hat\Psi^*(\vec\rho)\hat\Psi(\vec\rho) -n(\vec\rho)$ 
and this is the density fluctuations of the system.
The compressibility and number fluctuations have been measured in BEC 
experiments \cite{chicago,jila}.
We present the compressibility for a trapped dipolar gas in 
Fig. \ref{compress} (b) at $T/T_0=0.5$ for 2000 (blue), 3000 (red), 
and 4000 (black) in the $x$ (solid) and $y$ (dashed).
These results show that the density fluctuations roughly follow
the anisotropic density profile. Another way to state this is: that depending on
the local value of the density, the compressibility is determined almost uniquely.

In this work, we found that the correlations are anisotropic in a q2D
DBEC with a polarization axis tilted into the plane of motion.  
We studied the DBEC with the HFBP at at temperature of $T/T_C\sim0.5$.
At this temperature, quasiparticles are thermally
populated and total density is nearly isotropic
in the central region of the trap.
We computed the $g_1(\vec\rho;0)$ and $g_2(\vec\rho;0)$ correlations functions,
which have anisotropic character in the center of the gas
even though the density is isotropic in the central region of the DBEC.
This has measurable implications for interference properties and 
density fluctuations of the gas. 
We found that the anisotropic correlations occur due to roton like 
quasiparticle modes that can lead to local collapse along the polarization axis.

An immediate extension of this work is to study the static structure factor
such as was recently done in Ref. \cite{blair} at zero temperature. 
An important difference is that the static structure will become anisotropic, 
i.e. $S(k_x,k_y)$, and with present methods, it could be done at finite temperature \cite{stemp}.
This also presents a promising method to observe the anisotropic coherences predicted here.

Future work will be to investigate the impact of correlations 
at higher temperatures and strongly interactions
where they might impact the 
BKT transition \cite{BKT,BKTrev,dBKT}. This will require an improvement to the
method \cite{mora} or new approach altogether.  
Either way, it will be intriguing to study the
impact of the anisotropic dipolar interaction on vortex correlations near 
the BKT transition and the time dependence of such a gas.
The present work was not able to measure the decay of the correlation 
functions due to the relatively weak interaction strength or small number
the numerical method could handle.
If the next method can probe stronger interactions, 
we would hope to observe a different character in the decay of the correlation
functions along and perpendicular to the polarization axis.

{\it Acknowledgments,} the author is grateful for support from the
Advanced Simulation and Computing Program (ASC) and LANL which is operated 
by LANS, LLC for the NNSA of the U.S. DOE under Contract No. DE-AC52-06NA25396.
The author is grateful for discussion with A. Sykes, R. Behunin, and L. A. Collins.

\bibliographystyle{amsplain}

\end{document}